\documentclass[prl,reprint,amssymb,amsmath,superscriptaddress,nobibnotes]{revtex4-1}

\usepackage{graphicx}
\usepackage{color}
\usepackage{hyperref}
\usepackage{bm}
\usepackage{siunitx}

\allowdisplaybreaks

\newcommand{\umlaut}{\"}
\newcommand{\parag}{\\\indent}

\newcommand{\dsigma}{2D_s\tau_s+\sigma_{\rm eff}^2}
\newcommand{\vh}{v_{\rm h}}
\newcommand{\vv}{v_{\rm v}}

\newcommand{\seff}{\sigma_{\rm eff}}
\newcommand{\spp}{\sigma_{\rm pp}}
\newcommand{\spr}{\sigma_{\rm pr}}

\definecolor{olive}{rgb}{0.000 ,0.5216 ,0.1373}
\definecolor{RoyalRed}{RGB}{138,0,33}
\definecolor{skyblue}{RGB}{0 ,109 ,255}
\definecolor{RoyalGreen}{RGB}{37, 102, 43}
\definecolor{emerald}{RGB}{64,127,127}
\definecolor{gbule}{RGB}{12,127,172}

\hypersetup{
    bookmarks=true,         
    bookmarksnumbered=true,
    unicode=false,          
    pdftoolbar=true,        
    pdfmenubar=true,        
    pdffitwindow=false,     
    pdfstartview={FitH},    
    pdfauthor={Daisuke IIZASA},     
    colorlinks=true,       
    linkcolor=RoyalRed,      
    citecolor = RoyalRed,          
    linkbordercolor={1 0 0},
    urlcolor = RoyalRed, 
}
\begin{document}

\title{Control of Spin Relaxation Anisotropy by Spin-Orbit-Coupled Diffusive Spin Motion}

\author{Daisuke Iizasa}
\affiliation{Department of Materials Science, Tohoku University, Sendai 980--8579, Japan}
\author{Asuka Aoki} 
\affiliation{Department of Materials Science, Tohoku University, Sendai 980--8579, Japan}
\author{Takahito Saito} 
\affiliation{Department of Materials Science, Tohoku University, Sendai 980--8579, Japan}
\author{Junsaku Nitta}
\affiliation{Department of Materials Science, Tohoku University, Sendai 980--8579, Japan}
\affiliation{Center for Spintronics Research Network, Tohoku University, Sendai 980--8577, Japan}
\affiliation{Center for Science and Innovation in Spintronics (Core Research Cluster), Tohoku University, Sendai 980--8577, Japan}
\author{Gian Salis} 
\affiliation{IBM Research-Zurich, S\umlaut{a}umerstrasse 4, 8803 R\umlaut{u}schlikon, Switzerland.}
\author{Makoto Kohda} 
\affiliation{Department of Materials Science, Tohoku University, Sendai 980--8579, Japan}
\affiliation{Center for Spintronics Research Network, Tohoku University, Sendai 980--8577, Japan}
\affiliation{Center for Science and Innovation in Spintronics (Core Research Cluster), Tohoku University, Sendai 980--8577, Japan}
\affiliation{Division for the Establishment of Frontier Sciences, Tohoku University, Sendai 980-8577, Japan}

\begin{abstract}
Spatiotemporal spin dynamics under spin-orbit interaction is investigated in a (001) GaAs two-dimensional electron gas using magneto-optical Kerr rotation microscopy.
Spin polarized electrons are diffused away from the excited position, resulting in spin precession because of the diffusion-induced spin-orbit field.
Near the cancellation between spin-orbit field and external magnetic field, the induced spin precession frequency depends nonlinearly on the diffusion velocity, which is unexpected from the conventional linear relation between the spin-orbit field and the electron velocity.
This behavior originates from an enhancement of the spin relaxation anisotropy by the electron velocity perpendicular to the diffused direction.
We demonstrate that the spin relaxation anisotropy, which has been regarded as a material constant, can be controlled via diffusive electron motion.
\end{abstract}

\date{\today}

\maketitle

\begin{figure}
        \centering   
        \includegraphics[keepaspectratio, scale=0.5]{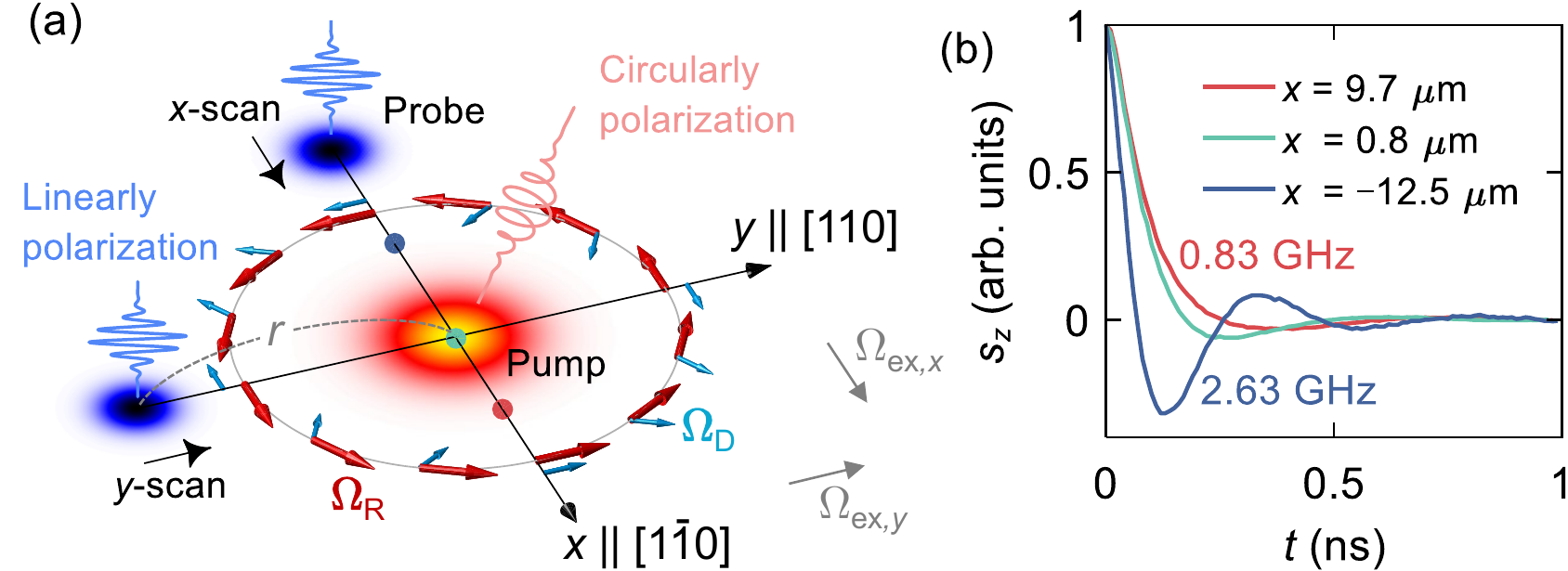}    
        \caption{(a) Sketch of a pump-probe scanning Kerr microscopy setup with Rashba $(\bm{\Omega}_{\rm R})$ and Dresselhaus ($\bm{\Omega}_{\rm D}$) fields as precession vectors. An external magnetic field is depicted as $\bm{\Omega}_{{\rm ex},x}$ and $\bm{\Omega}_{{\rm ex},y}$ as a precession vector for $y$- and $x$-scans configurations, respectively. A circularly polarized pump beam excites a spin polarization $s_z$. A linearly polarized probe beam detects $s_z$ at a delay time $t$ and a position $(x,y)$. (b) Measured $s_z$ at different $x$ positions highlighted as colored circles in (a).}
\label{one}
\end{figure}

Precise control of spin motion is a prerequisite from fundamental physics to spintronics and quantum information technology \cite{Wolf2001,Awschalom2007,BehinAein2010,Waldrop2016}. In a semiconductor quantum well (QW), Rashba \cite{Rashba1960,Bychkov1984} and Dresselhaus \cite{Dresselhaus1955} spin--orbit (SO) interactions act as effective magnetic fields for moving electrons, enabling coherent spin control via precession, whereas spin relaxation occurs simultaneously because of an interplay between the SO field and the random motion of electrons \cite{Dyakonov1971}. Both spin precession and relaxation processes are closely tied to one another solely by SO interaction \cite{Nitta1997}. For stationary electrons with mean zero velocity, the correlation between precession and relaxation triggers a modulation of spin precessional motion, known as spin relaxation anisotropy \cite{Averkiev1999,Averkiev2002,Kainz2003,Dohrmann2004,Morita2005,Averkiev2006,Schreiber2007,Stich2007,Larionov2008,Griesbeck2012}. For spin rotation by external and/or SO fields in a QW, spins along growth and in-plane orientations do not experience identical torques because of the in-plane orientation of the SO fields. This situation induces anisotropic spin relaxation \cite{Averkiev1999,Averkiev2002,Kainz2003,Dohrmann2004,Morita2005,Averkiev2006,Schreiber2007,Stich2007,Larionov2008,Griesbeck2012} and modulates the spin precession frequency \cite{  Dohrmann2004,Morita2005,  Averkiev2006, Schreiber2007,Larionov2008, Griesbeck2012}. Because SO fields are well defined for stationary electrons, the spin relaxation anisotropy has been regarded as a material constant. However, for moving electrons with a finite net velocity induced by drift \cite{Kato2004,Meier2007, Studer2010,Walser2012b,Altmann2016,Saito2019} and diffusion \cite{ Altmann2016, Saito2019, Kohda2015, Kawaguchi2019}, the electron trajectory further modulates SO fields and directly affects the spin relaxation anisotropy through the momentum-dependent spin precession. Moreover, the spin relaxation anisotropy is not limited to particular materials such as III--V semiconductors because the anisotropic SO fields are ubiquitous in solid states, with spin-momentum locking in topological insulators \cite{Hsieh2009,Hasan2010}, Rashba interface in oxides \cite{Caviglia2009}, metal interfaces \cite{Meier2008}, and Zeeman-type SO field in 2D materials \cite{Yuan2013}. Consequently, unveiling the effects of moving electrons on the modulation of spin relaxation anisotropy and induced precession frequency are expected to be crucially important for future spintronics, topological electronics, and quantum information technologies. Despite this, earlier studies of spin relaxation anisotropy have remained limited only to stationary cases \cite{Dohrmann2004,Morita2005,  Averkiev2006, Schreiber2007,Larionov2008, Griesbeck2012}.
\parag
Here, we experimentally manifest control of spin precessional motion via spin relaxation anisotropy by diffusive spin motion in a GaAs-based QW. When the SO field under diffusive motion is nearly compensated by a constant external magnetic field, the spin precession frequency is no longer linear to the diffusion velocity. This behavior cannot be anticipated from a conventional spin drift/diffusion model. It is explained by a modulation of the spin relaxation anisotropy. The evaluated spin relaxation anisotropy, which exhibits six-fold enhancement from the stationary case, is explained by a tilting of the spin precession axis from the direction of external magnetic field caused by the electron diffusive motion. We influence the spin relaxation anisotropy, as reported for the first time, by precisely controlling the electron motion.
\parag
The structure examined for this study was an n-doped 20-nm-thick (001) GaAs QW. In this system, we obtain SO fields characterized by the Rashba parameter $\alpha\ (<0)$, the Dresselhaus parameter $\beta=\beta_1-\beta_3\ (>0)$, with linear $\beta_1=-\gamma\langle k_z^2\rangle$ and cubic term $\beta_3=-\gamma k_{\rm F}^2/4$. Here, $\langle k_z^2\rangle$ denotes the expected value of the squared wavenumber in the QW. The bulk Dresselhaus coefficient is $\gamma<0$. The Fermi wavenumber is $k_{\rm F}=\sqrt{2\pi n_s}$. The carrier density and mobility measured using a Hall bar device were $n_s=\SI{1.72e11}{cm^{-2}}$ and $\SI{11.2e4}{cm^2 V^{-1} s^{-1}}$, respectively, at 4.2 K. To detect the diffusive spin dynamics, spatiotemporal Kerr rotation microscopy is performed using a mode-locked Ti:sapphire laser emitting 2-ps-long pulses at a 79.2 MHz repetition rate. Figure \ref{one}(a) depicts an experimental configuration for pump and probe beams with Rashba and Dresselhaus SO fields. Therein, $\bm{\Omega}_{\rm R}$ and $\bm{\Omega}_{\rm D}$ respectively represent the spin precession frequency vectors. A circularly polarized pump beam with Gaussian sigma-width $\sigma_{\rm pp}$ is focused onto the sample surface to excite spin polarization $s_z$ along the growth direction. A linearly polarized probe beam (spot size $\sigma_{\rm pr}$) detects $s_z$ at delay time $t$ and arbitrary position by a motor-controlled scanning mirror. All optical measurements are taken at 30 K.


\begin{figure}[t]
        \centering   
        \includegraphics[keepaspectratio, scale=0.5]{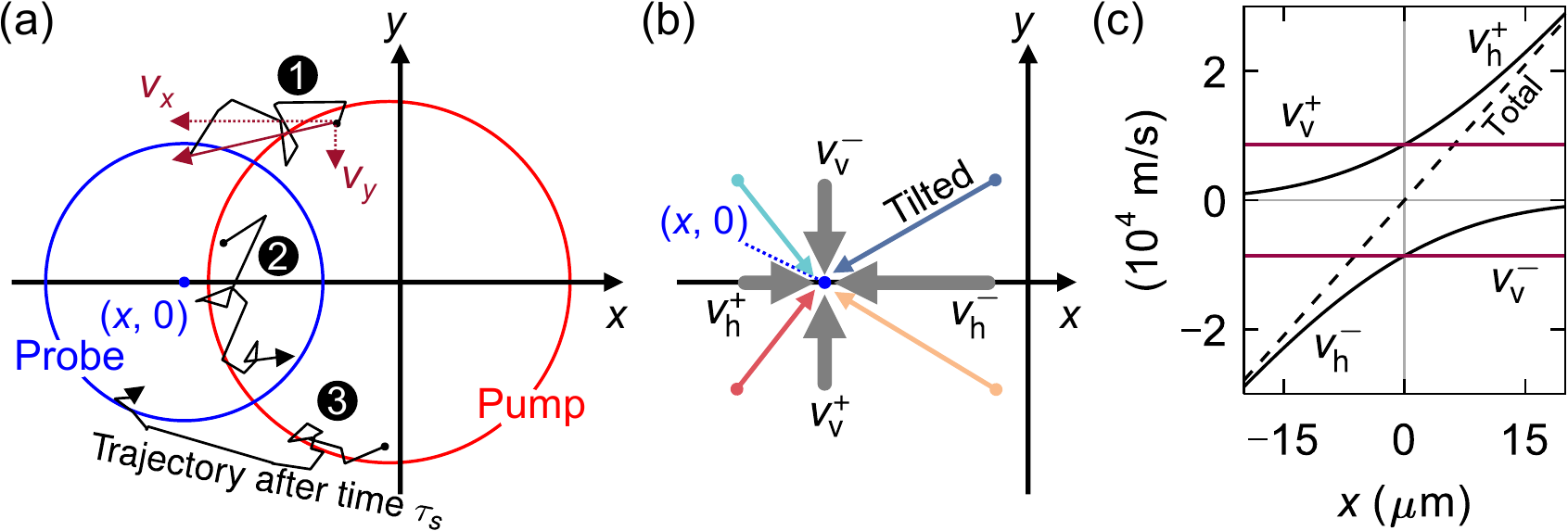}    
        \caption{(a) $x$-scan configuration where the probe center is separated by a vector $(x, 0)$ from the pump center $(0, 0)$. Electron trajectories exist with an average velocity that is tilted from the $x$-axis, contrary to the macroscopic diffusion velocity [Eq.~(\ref{difvelo})]. (b) Mean velocity components of horizontal and vertical directions, $\vh^\pm$ and $\vv^\pm$. (c) Calculated $\vh^\pm$, $\vv^\pm$ based on Eqs.~(\ref{vh}) and (\ref{vv}). $\vv^\pm$ are constant with respect to $r$ here, whereas $\vh^\pm$ depends on the position.}
\label{three}
\end{figure}

The spin precession frequency induced by a velocity $\bm{v} =(v_x,v_y)$ in an external magnetic field $\bm{B}_{\rm ex}  =(B_x,B_y)$ is generally described as
\begin{align}
\Omega_{x,y} (v_{y,x})=\cfrac{2m}{\hbar^2} (\pm\alpha+\beta)v_{y,x}+\cfrac{g\mu_{\rm B}}{\hbar} B_{x,y}. \label{prevec}
\end{align}
Here $g<0$ stands for the electron $g$ factor, $\mu_{\rm B}$ denotes the Bohr magneton, $\hbar$ is the reduced Planck’s constant, and $m=0.067m_0$ expresses effective electron mass of GaAs. The diffusion velocity $v_{\rm dif}$, which is controlled by the center-to-center distance $r$ between pump and probe spots, is defined as
\begin{align}
v_{\rm dif}=2D_s r/(2D_s \tau_s+\sigma_{\rm eff}^2), \label{difvelo}
\end{align}
where $D_s$ is the spin diffusion constant, $\tau_s$ represents the D'yakonov-Perel' spin relaxation time, and the convoluted spot size $\seff$ is defined by $\seff^2=\spp^2+\spr^2$ \cite{Kohda2015}. Also, $\tau_s$ is a result of the replacement of $t=\tau_s$ because our system satisfies $2D_s \tau_s\ll\seff^2$ and small $\tau_s$. By changing the probe position along the $x$-axis ($y$-axis) \cite{Kohda2015}, i.e., the distance $r$ in Eq.~(\ref{difvelo}) , one can set the diffusion velocity $v_{\rm dif} =v_x \ (v_{\rm dif} =v_y)$ and thereby modulate the spin precession frequency [$\Omega_y (v_x)$ or $\Omega_x (v_y)$ in Eq.~(\ref{difvelo})]. Figure \ref{one}(b) shows the time evolution of the experimental Kerr signal ($s_z$) at different probe positions ($x=9.7,0.8$ and $-12.5$ $\mu$m) in an $x$-scan ($\seff=8.1$ $\mu$m and $B_y=+0.45$ T). The spin precession frequency depends strongly on the probe position, reflecting the momentum (velocity) dependent SO field induced by the finite diffusion velocity. We systematically measured Kerr signals with different positions on the $x$- and $y$-axes with several spot sizes $\seff$. We extracted the precession frequency $|\Omega_{\rm meas} |$ by fitting the normalized Kerr signal $s_z=\exp⁡(-t/\tau_s)\cos⁡(2\pi|\Omega_{\rm meas} |t+\phi)$ with phase shift $\phi$.
\parag
\begin{figure}
        \centering   
        \includegraphics[keepaspectratio, scale=0.5]{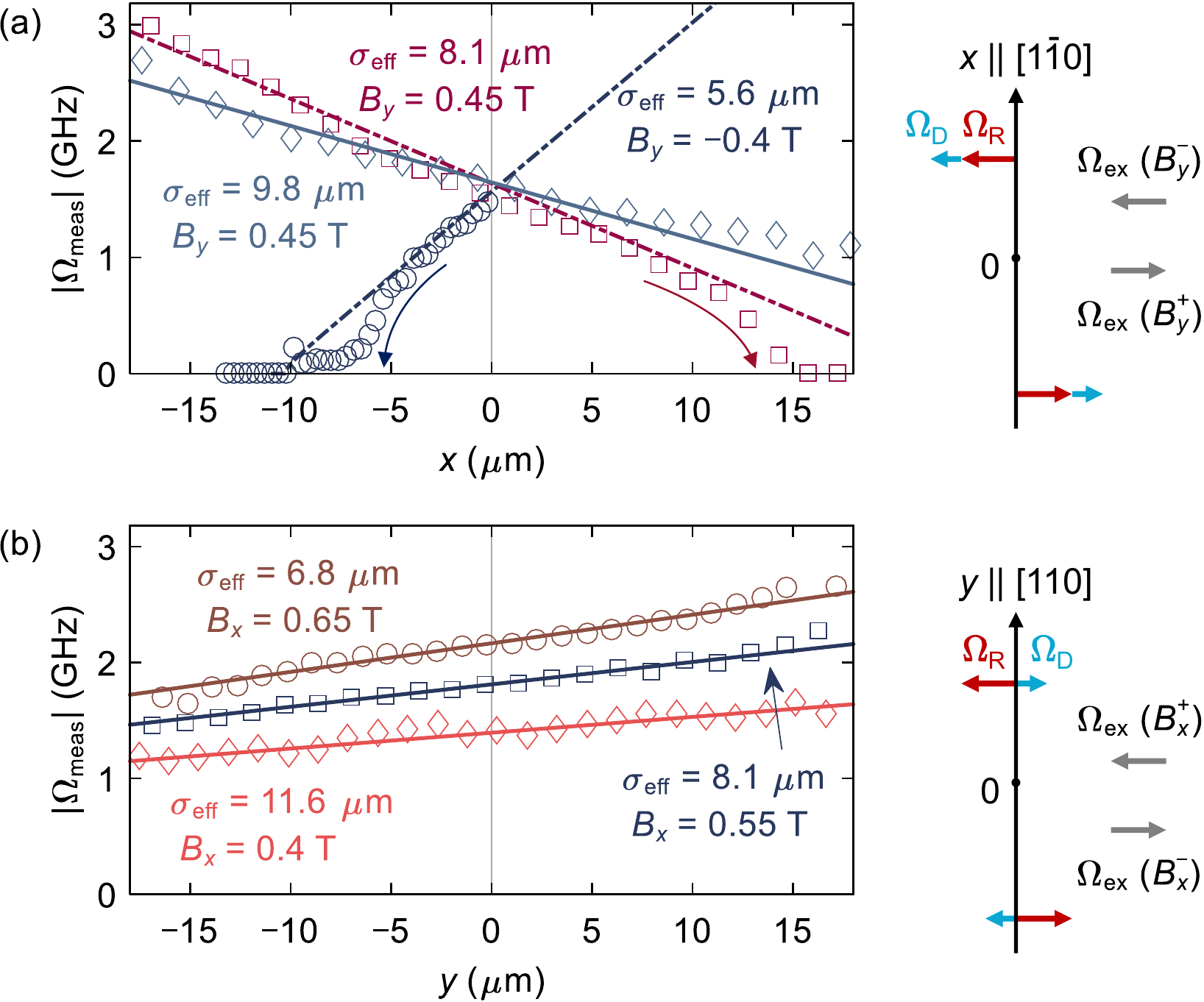}    
        \caption{Measured spin precession frequency $|\Omega_{\rm meas} |$ obtained for different $\seff$ and $\bm{B}_{\rm ex}$ and for scans of the pump-probe separation along $x$ (a) and $y$ (b). Diffusing spins experience strong SO fields for $x$-scan, but weak fields for $y$-scan. All symbols represent experimental data. All solid lines show linear fits. Dashed lines in (a) correspond to the nonlinear fits based on Eq.~(\ref{nonlinear}) with $\Gamma_{\rm at}=-0.076$ GHz.}
\label{two}
\end{figure}
Figures~\ref{two}(a) and \ref{two}(b) summarize extracted $|\Omega_{\rm meas} |$ in $x$- and $y$-scans. For the $y$-scan [Fig.~\ref{two}(b)], $|\Omega_{\rm meas} |$ varies linearly with the $y$ position for all conditions of $B_x$ and $\seff$, reflecting the linear dependence of $v_{\rm dif}$ on the $y$ position, as presented in Eq.~(\ref{difvelo}). In addition, when $\seff$ decreases from 11.6 to 6.8 $\mu$m, the slope $d\Omega_{\rm meas}/dy$ increases gradually, which agrees well with Eq.~(\ref{difvelo}) and which is consistent with earlier reports of the literature \cite{Kato2004, Studer2010, Walser2012b, Saito2019,Kohda2015,Kawaguchi2019}. For the $x$-scan [Fig.~\ref{two}(a)], however, a linear variation of $|\Omega_{\rm meas} |$ on the $x$ position is only observed for $\seff=9.8$ $\mu$m and $B_y=+0.45$ T (diamond symbols). Reducing $\seff$ to 8.1 and 5.6 $\mu$m exhibits a deviation from a linear variation; notably most pronounced when $|\Omega_{\rm meas} |$ approaches zero. This cannot be explained using the conventional linear relation between electron velocity and SO field. To understand this effect, we first evaluate the SO parameters from the linear frequency variation. From linear fits depicted as solid lines in Figs.~\ref{two}(a) and \ref{two}(b), we obtain $\alpha=\SI{-2.89e-13}{eVm}$, $\beta_1=\SI{1.86e-13}{eVm}$, and $\beta_3=\SI{0.22e-13}{eVm}$. Also, $g=-0.268$ is estimated at $r=0\   (v_{\rm dif}=0)$. We assume $g<0$ based on the QW thickness \cite{Walser2012b}. Also, $D_s=\SI{0.0195}{m ^2/s}$ is derived from the measured $\tau_s=\SI{75}{ps}$ at $\bm{B}_{\rm ex}=\bm{0}$ T \cite{Henn2016}. Using evaluated $\beta_1,\beta_3$, and $n_s$, we obtain $\gamma=\SI{-8.31}{eV}$\AA$^3$ which is consistent with values reported in the literature \cite{Kohda2017}. To explain our observation, we introduce in analogy to anisotropic spin relaxation for stationary electrons modified spin precession frequencies~\cite{Dohrmann2004,Morita2005,  Averkiev2006, Schreiber2007,Larionov2008, Griesbeck2012},
\begin{align}
\Omega_x^\ast=\sqrt{\Omega_x (v_{\rm dif})^2-\Gamma_{\rm at}^2},\ \Omega_y^\ast=\sqrt{\Omega_y (v_{\rm dif})^2-\Gamma_{\rm at}^2}, \label{nonlinear}
\end{align}
where the anisotropic term~\cite{Averkiev2006, Larionov2008} is
\begin{align}
\Gamma_{\rm at}(\Theta)=-(\Gamma_x \cos^2⁡\Theta+\Gamma_y \sin^2⁡\Theta)/2.   \label{anisotropy}
\end{align}
Here the relaxation rate of spins oriented along $x$- and $y$-axes is $\Gamma_{x,y}=(4D_s m^2/\hbar^4)[(\mp\alpha+\beta)^2+\beta_3^2]$, respectively, and $\Theta\in[0,2\pi]$ is the direction of the spin precession axis, defined as in-plane polar angle from $+x$- toward $+y$-axis. The term $\Gamma_{\rm at}(\Theta)$ describes the relaxation anisotropy between the two relevant orthogonal crystal axes and is responsible for a correction of the precession frequency [Eq.~(\ref{nonlinear})]. For the $y$-scan ($\bm{B}_{\rm ex}=(B_x,0)$) spins precess in the $y$-$z$ plane, and $\Gamma_{\rm at} (\Theta=0)=-\Gamma_x/2=(\Gamma_y-\Gamma_z)/2$ denotes half of the difference of the relaxation rate between $y$- and $z$-axes, where $\Gamma_z=\Gamma_x+\Gamma_y$ is the relaxation rate along the $z$-axis. For the $x$-scan ($\bm{B}_{\rm ex}=(0,B_y)$), $\Gamma_{\rm at} (\pm\pi/2)=-\Gamma_y/2$. Because $\Gamma_{\rm at} (\Theta)$ additionally contributes to the spin precession frequency shown in Eq.~(\ref{nonlinear}), $\Omega_{x,y}^\ast$ shows a nonlinear dependence on the probe position $r$, which becomes pronounced when the precession frequency induced by external and SO fields becomes comparable to $\Gamma_{x,y}/2$. Based on the experimentally evaluated values for $\alpha,\beta_1,\beta_3$, and $D_s$, we calculate $-\Gamma_y/2=-0.076$ and $-\Gamma_x/2=-0.99$ GHz.  For $\seff=5.6$ and 8.1 $\mu$m, the calculated $\Omega_y^\ast$ are shown as dashed lines in Fig.~\ref{two}(a). The calculated values only reproduce the experimental data in a linear frequency region. The rapid decrease of $\Omega_y^\ast$ that occurs below 0.8 GHz cannot be explained by $-\Gamma_y/2=-0.076$ GHz.
\parag
According to Eq.~(\ref{anisotropy}), the frequency modulation caused by the relaxation anisotropy depends on the direction of the precession axis ($\Theta$). For stationary electrons under $\bm{B}_{\rm ex}$, where the SO field does not contribute to frequency modulation, $\Theta$ is well-defined by the direction of $\bm{B}_{\rm ex}$. However, for moving electrons, the precession axis is defined by the sum of $\bm{B}_{\rm ex}$ and the SO field, implying that the electron trajectories under diffusion further modulate $\Theta$. Because various electron trajectories can lead from the pump spot to the probe spot, the precession axis is no longer well defined by $\bm{B}_{\rm ex}$ because of different diffusion velocity vectors $\bm{v} =(v_x,v_y)$ in time $\tau_s$ [different arrows in Fig.~\ref{three}(a)]. Specifically examining one single trajectory with average velocity $\bm{v}$, its direction of average precession axis $\Theta(v_x,v_y)=\arctan(\Omega_y (v_x)/\Omega_x (v_y))$ can be obtained from Eq.~(\ref{prevec}). Entering $\Theta(v_x,v_y)$ into Eq.~(\ref{anisotropy}) directly reveals the velocity-dependent form of the anisotropic term
\begin{align}
\Gamma_{\rm at} (v_x,v_y)=-\cfrac{\Gamma_x \Omega_x (v_y )^2+\Gamma_y \Omega_y (v_x )^2}{2\left(\Omega_x (v_y )^2+\Omega_y (v_x )^2\right)}.  \label{vaniso}
\end{align}
It is noteworthy that the precession frequencies caused by opposite velocities ($\pm v_{y,x}$) do not cancel each other because $\Omega_{x,y}$ enter as squares. This finding contrasts to the notion of a single (averaged) diffusion velocity for given pump and probe spots [described by Eq.~(\ref{difvelo})], corresponding to the mean value of all velocity vectors leading from the pump to the probe spots [Fig.~\ref{three}(a)]. Actually, Eq.~(\ref{vaniso}) rather suggests that the anisotropic term is defined by the microscopic behavior of the electron motion (velocity). Therefore, we sort all diffusion velocity vectors along the $x$- and $y$-axes according to their sign and evaluate the mean values of horizontal and vertical velocities at probe positions $\vh^\pm$ and $\vv^\pm$, respectively, as indicated by silver bold arrows in Fig.~\ref{three}(b). Each velocity is described as
\begin{align}
\vh^\pm &=\pm\sigma_{\rm eff}\sqrt{\cfrac{\Sigma}{\pi \tau_s}}\,e^{  -\frac{r^2\Sigma \tau_s}{\sigma_{\rm eff}^2}}+\Sigma r\,\mathrm{erfc}\left( \mp\cfrac{r\sqrt{\Sigma \tau_s}}{\sigma_{\rm eff}} \right), \label{vh}\\
\vv^\pm&=\pm\sigma_{\rm eff}\sqrt{{\Sigma}/({\pi \tau_s})},\label{vv}
\end{align}
where $\Sigma=D_s/(\dsigma)$. The complementary error function is denoted by erfc. The $\pm$ signs in $\vh^\pm$  and $\vv^\pm$ respectively correspond to the positive/negative velocity components along the horizontal or vertical axes. Figure~\ref{three}(c) shows calculated $\vh^\pm$  and $\vv^\pm$. The $\vh^+$ increases rapidly in the $+x$ region, whereas $\vh^-$ has the opposite tendency because of radial diffusion of electrons from the excited pump spot.
\parag
When the probe spot is displaced along the $-x$-axis [Fig.~\ref{three}(a)], the average velocity vector points to the $-x$-axis because there, $|\vh^- |>|\vh^+ |$. Remarkably, vertical velocities $\vv^\pm$ are independent of probe position and are of similar size as $\vh^\pm$. This suggests that the tilted velocity vectors should contribute to $\Gamma_{\rm at}$ and the precession axis ($\Theta$) no longer points along $\bm{B}_{\rm ex}$. For the total mean velocity $\vh^++\vh^-+\vv^++\vv^-$ as sketched in Fig.~\ref{three}(b), the $\vv^\pm$ cancel out each other and are linear in the position [dashed line in Fig.~\ref{three}(c)], corresponding to the conventional macroscopic diffusion velocity [Eq.~(\ref{difvelo})].
\parag
\begin{figure}[t]
        \centering   
        \includegraphics[keepaspectratio, scale=0.5]{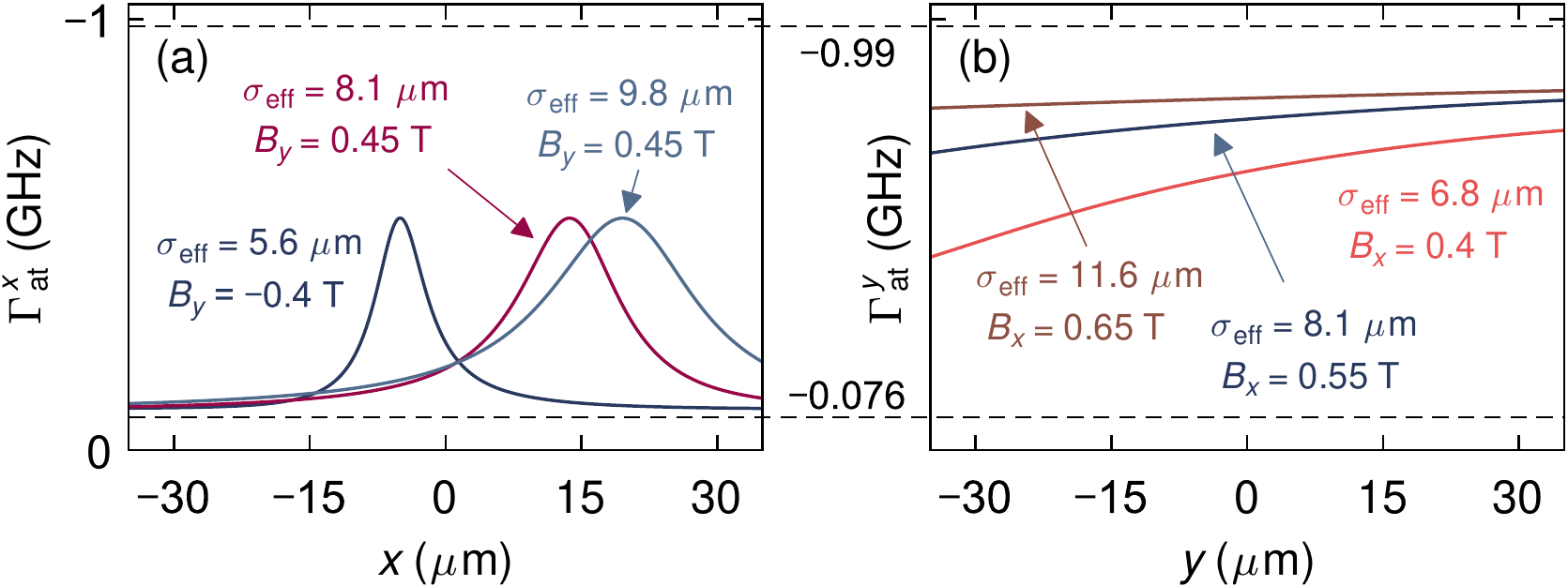}    
        \caption{(a) Calculated anisotropy term $\Gamma_{\rm at}^{x,y}$ for $x$-scan (a) and $y$-scan (b) as obtained from using Eq.~(\ref{final}). Both $\Gamma_{\rm at}^{x,y}$ are modulated by $r$. Parameters for the calculation are all experimentally determined values.}
\label{four}
\end{figure}
By considering both horizontal and vertical velocities [Eqs.~(\ref{vh}) and (\ref{vv})] depicted in Fig.~\ref{three}(b), we average over all contributions to the anisotropic term [Eq.~(\ref{vaniso})] for the $x$-scan with
\begin{align}
\Gamma_{\rm at}^x=&\big[\Gamma_{\rm at} (\vh^+,\vv^+)+\Gamma_{\rm at} (\vh^+,\vv^-)\nonumber\\
&+\Gamma_{\rm at} (\vh^-,\vv^+)+\Gamma_{\rm at} (\vh^-,\vv^-)\big]/4.   \label{final}
\end{align}
For the $y$-scan configuration, $\Gamma_{\rm at}^y$ is obtained from Eq.~(\ref{final}) by flipping $\vh$ and $\vv$ with each other. We calculate $\Gamma_{\rm at}^{x,y}$ in Figs.~\ref{four}(a) and \ref{four}(b) with parameters evaluated from the experimental conditions. The $\Gamma_{\rm at}^x$ exhibits a peak structure corresponding to the cancellation between external and SO fields, i.e. $\Omega_y (v_{\rm dif})=0$. At this position, $\Gamma_{\rm at}^x$ is enhanced by more than six times from $-\Gamma_y/2=-0.076$ GHz. Such a peak structure is observed consistently for different spot sizes and $B_y$ values. The enhanced $\Gamma_{\rm at}^x$ is the consequence of a tilting of the spin precession axis away from $\bm{B}_{\rm ex}$ direction due to the $\vv^\pm$ components that introduce a precession contribution $\Omega_x (\vv^\pm )^2$. As seen from Eq.~(\ref{vaniso}),  $\Gamma_x \Omega_x (\vv^\pm )^2$ is introduced in the numerator of the expression for $\Gamma_{\rm at}$. For $y$-scan, $\Gamma_{\rm at}^y$ is only gently modulated with $y$ because, in this case, the additional contribution proportional to $\Gamma_y \Omega_y (\vv^\pm)$ is weak compared to the case of $x$-scan (because $\Gamma_x \gg \Gamma_y$). In other words, a small SO field along the $y$-axis does not tilt the spin precession axis significantly. In both $x$- and $y$-scans, when the magnitude of $\bm{B}_{\rm ex}$ becomes sufficiently large compared to the SO field, $\Gamma_{\rm at}^{x,y}$ converges respectively to the stationary cases of $-0.076$ GHz and $-0.99$ GHz.
\parag
\begin{figure}[t]
        \centering   
        \includegraphics[keepaspectratio, scale=0.5]{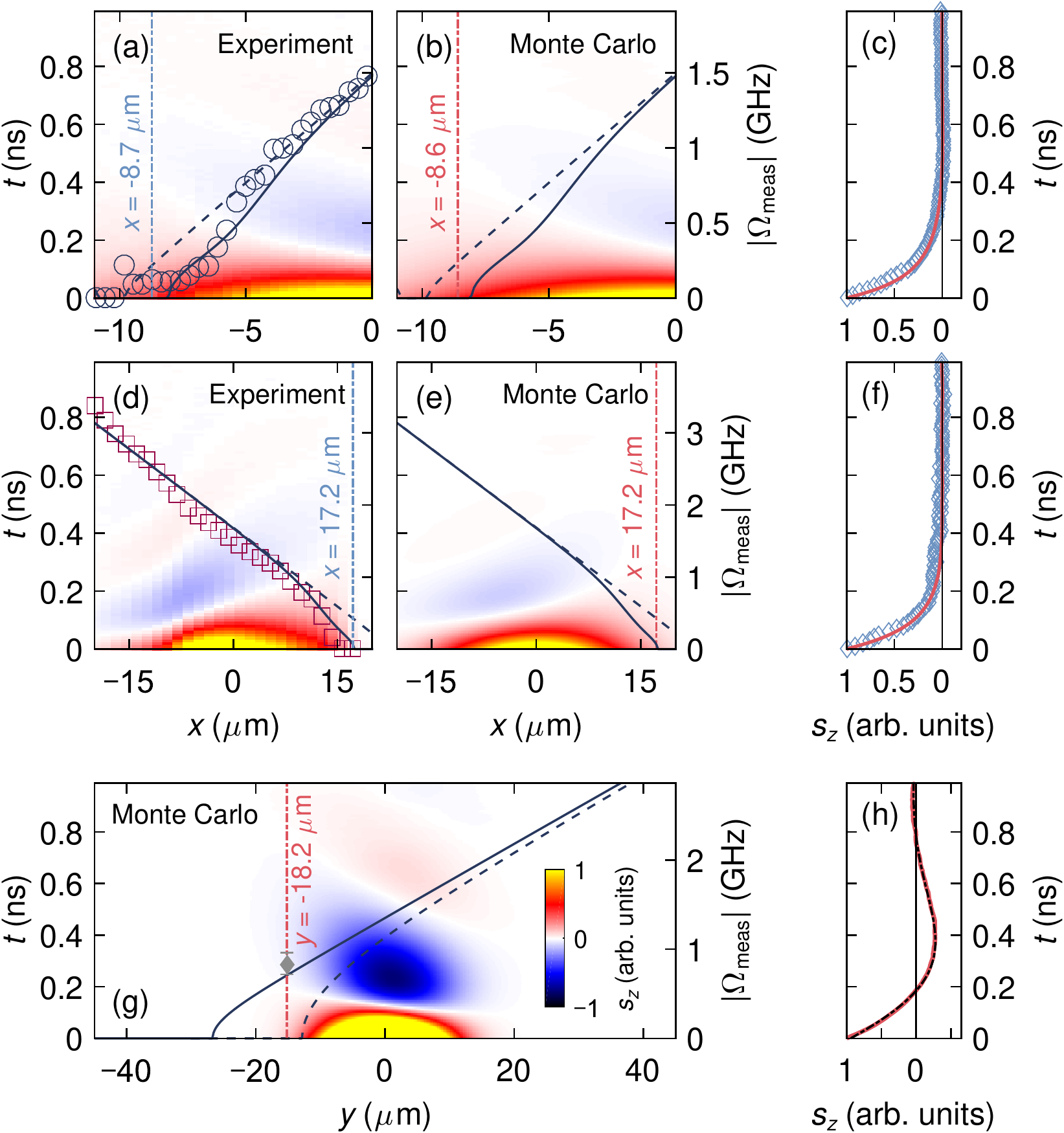}    
        \caption{Experimental and Monte Carlo simulated time–space records of $s_z$ in $x$-scan for (a) and (b) with $\seff=5.6$ $\mu$m at $B_y=-0.4$ T, and for (d) and (e) with $\seff=8.1$ $\mu$m at $B_y=0.45$ T. All solid lines are $\Omega_y^\ast$ calculated with new $\Gamma_{\rm at}^x$, and dashed lines with conventional anisotropic term $-\Gamma_y/2=-0.076$ GHz. The time evolution of $s_z$ is shown at (c) for $x\approx-8.6$ $\mu$m and in (f) for $x=-17.2$ $\mu$m for experimental data (diamonds) and for Monte Carlo simulation results (red solid lines). (g) Monte Carlo simulated time-space records of $s_z$ in a $y$-scan with $\seff=5.6$ $\mu$m, $B_x=0.4$ T and $g=-0.268$: the solid line shows $\Omega_x^\ast$ with new $\Gamma_{\rm at}^y$; the dashed line is obtained with conventional value $-\Gamma_x/2=-0.99$ GHz. The gray diamond is the extracted frequency from the (h) time evolution of $s_z$ at $y=-18.2$ μm with Monte Carlo simulation (red solid) and the fitted curve (dotted black).}
\label{five}
\end{figure}
Finally, we reproduce the nonlinear behavior of the $x$-scan quantitatively by considering the calculated $\Gamma_{\rm at}^{x,y}$ in Fig.~\ref{four}. Figure~\ref{five}(a) shows the experimentally obtained results of color-coded $s_z$ in an $x$-scan and a frequency analysis (circles) for $\seff=5.6$ $\mu$m with $B_y=\SI{-0.4}{T}$. Solid and dashed lines respectively correspond to the calculated $\Omega_y^\ast$ based on Eq.~(\ref{nonlinear}) with our new $\Gamma_{\rm at}^x$ [Eq.~(\ref{final})] and the conventional $-\Gamma_y/2=\SI{-0.076}{GHz}$. The remaining parameters of Eq.~(\ref{nonlinear}) are all experimentally obtained. The $\Omega_y^\ast$ obtained with the new $\Gamma_{\rm at}^x$ shows excellent agreement with the experimental values, including the nonlinear variation. To confirm the enhancement of $\Gamma_{\rm at}$ further, we conducted Monte Carlo (MC) simulation. Figure~\ref{five}(b) presents the simulated Kerr traces, showing good agreement with the experimentally obtained result. At $x\approx -8.6$ $\mu$m, whereas $\Omega_y^\ast$ with conventional $-\Gamma_y/2$ suggests finite precession with 0.19 GHz, $\Omega_y^\ast$ with the new $\Gamma_{\rm at}^x$ shows a halt of spin precession. As presented in Fig.~\ref{five}(c), the time evolution of $s_z$ for both the experiment (diamond) and MC simulation (red solid) at $x\approx -8.6$ $\mu$m confirms a simple exponential decay with no oscillatory behavior. Similar results also hold for $\seff=8.1$ $\mu$m with $B_y=0.45$ T in Figs.~\ref{five}(d)--\ref{five}(f), supporting enhancement of the anisotropic term. We also compare $y$-scan by MC simulation for parameters $\seff=5.6$ $\mu$m and $B_x=0.4$ T [Fig.~\ref{five}(g)]. Solid and dashed lines are calculated values of $\Omega_x^\ast$ with new $\Gamma_{\rm at}^y$ and conventional $-\Gamma_x/2=-0.99$ GHz, respectively, where $\Omega_x^\ast$ is enhanced for negative $y$ values for new $\Gamma_{\rm at}^y$. This point is confirmed further in Fig.~\ref{five}(h) by plotting the time evolution of $s_z$ at $y=-18.2$ $\mu$m. The spin precession frequency obtained from MC simulation at $y=-18.2$ $\mu$m [grey diamond in Fig.~\ref{five}(g)] shows good agreement with our new model. The quantitative agreement shown above reveals clearly that precession by the relaxation anisotropy is not a material constant parameter, but is rather controlled by diffusive spin motion.
\parag
In conclusion, we have experimentally observed an enhancement of the spin relaxation anisotropy by diffusive spin motion. We measured the precession frequency in an external magnetic field by changing the relative distance between excited pump and detected probe positions in a spatiotemporal Kerr rotation microscope. Because of various electron trajectories for electrons travelling from the pump to the probe position, the spin precession axis is tilted substantially from the external magnetic field direction when the diffusion-induced SO field nearly compensates the magnitude of the external magnetic field. Such a SO-coupled spin-diffusive motion controls the relaxation anisotropy. It is detected as a nonlinear precession frequency modulation. Whereas the relaxation anisotropy is regarded as a constant parameter for stationary electrons, it becomes controllable for moving electrons. This effect also points out a threshold to start spin precession at a certain velocity. Because this effect is not limited only to diffusive motion, but also can be controlled by drift and ballistic transport, our findings link the effect of the precise control of spin states to future spintronics and quantum information technology.

\begin{acknowledgements}
We acknowledge financial support from the Japanese Ministry of Education, Culture, Sports, Science, and Technology (MEXT) Grant-in-Aid for Scientific Research (Nos. 15H02099, 15H05854, 25220604, and 15H05699), EPSRC-JSPS Core-to-Core program, and the Swiss National Science Foundation through the National Center of Competence in Research (NCCR) QSIT. D.I. thanks the Graduate Program of Spintronics at Tohoku University, Japan, for financial support.
\end{acknowledgements}

\bibliographystyle{apsrev4-1.bst}
\bibliography{bibspaceMaster}

\end{document}